\documentclass[leqno,draft,11pt]{article}
\usepackage{amssymb,amsmath,theorem,a4wide}
\usepackage[final]{graphicx}

\def\enumup{}

\ifx\mydefs\undefined\else \fi
\let\mydefs\relax

\allowdisplaybreaks[2]

\ifx\loadcyr\undefined\else
\input cyracc.def
\makeatletter
 at 1\@ptsize pt
 at 1\@ptsize pt
 at 1\@ptsize pt
\makeatother

\fi

\let\M\mathit
\def\gobble#1{}
\def\fixsup#1#2{{#1\let\dp\gobble\mathstrut}^#2_}

\def\bme{\hskip.75em\relax}

\def\iff{\quad\text{iff}\quad}
\let\LOR\bigvee%\limits ?
\let\ET\bigwedge%\limits ?

\def\?{\mathbin?}

\newbox\circlebox
\setbox\circlebox\hbox{$\bigcirc$}
\def\circled#1{%
  \setbox0\hbox to\wd\circlebox{\hss$#1$\hss}\wd0=0pt
  \box0\copy\circlebox}

\let\fii\varphi
\let\tet\vartheta
\let\ep\varepsilon

\def\greek#1{$\expandafter\greeknum\csname c@#1\endcsname$}
\makeatletter
\def\greeknum#1{\ifcase#1\or\alpha\or\beta\or\gamma\or\delta\or\ep
      \or\digamma\or\zeta\or\eta\or\tet\or\iota\else\@ctrerr\fi}
\makeatother

\def\p#1{\langle#1\rangle}

\def\lh#1{\lvert#1\rvert}

\let\bez\smallsetminus
\let\sdif\vartriangle
\let\sset\subseteq
\let\nsset\nsubseteq
\let\Sset\supseteq

\let\onto\twoheadrightarrow

\def\pw#1{\mathcal P(#1)}
\let\nul\varnothing

\def\twoprimes{\raise.2\fontdimen6\scriptfont2\hbox{$\scriptstyle\prime\prime$}}
\newcommand\rpair[3][3em]{\mathrel{%
   \begin{matrix}%
     \strut\smash{\xrightonto{\hbox to#1{\hss$#2$\hss}}}\\[-1.7ex]%
     \strut\smash{\xleftembed[\hbox to#1{\hss$#3$\hss}]{}}%
   \end{matrix}}}
\makeatletter
\newcommand\xrightonto[2][]{\ext@arrow 0359\rightontofill{#1}{#2}}
\newcommand\xleftembed[2][]{\ext@arrow 3095\leftembedfill{#1}{#2}}
\def\leftembedfill{\arrowfill@\leftarrow\relbar\hookleftnoarrow}
\def\rightontofill{\arrowfill@\relbar\relbar\onto}
\def\hookleftnoarrow{\DOTSB\relbar\joinrel\rhook}
\makeatother

\mathchardef\#="2023 % \mathbin\#

\ifx\busspf\undefined\else
\usepackage{bussproofs}
\EnableBpAbbreviations

\fi

\let\dia\Diamond

\ifx\symlasy\undefined

  \def\centerdot#1{%
    \setbox0\hbox{$\mathop{#1}$}\dimen0 \ht0
    \setbox0\hbox{$#1$}\advance\dimen0 -\ht0
    \setbox2\hbox to\wd0{\hss$\mathop{\cdot}$\hss}\wd2=0pt
    \lower\dimen0\box2\box0 }
\else
  \def\centerdot#1{%
     \setbox0\hbox{$#1$}%
     \raise0.206\ht0\hbox to\wd0{\hss$\cdot$\hss}%
     \kern-\wd0 \box0 }
\fi

\def\adm{\mathrel{|\mkern-8.5mu\sim}}
\def\nadm{\mathrel{|\mkern-8.5mu\not\sim}}
\def\ru{\mathrel/}

\let\sls|

\def\Up{{\setbox0\hbox{$\uparrow$}%
         \lower\dp0\hbox to\wd0{\hss\vrule width4pt height.4pt\hss}%
         \kern-\wd0\box0}}
\def\UP{{\setbox0\hbox{$\uparrow$}%
         \lower\dp0\hbox to\wd0{\hss\vrule width4pt height.4pt\hss}%
         \kern-\wd0\copy0\kern-\wd0\raise.35ex\box0}}
\def\Down{{\setbox0\hbox{$\downarrow$}%
         \raise\ht0\hbox to\wd0{\hss\vrule width4pt depth.4pt\hss}%
         \kern-\wd0\box0}}

\DeclareMathOperator\Int{int}%\int clashes
\DeclareMathOperator\chull{C}

\let\cxt\mathrm
\def\np{\cxt{NP}}
\def\conp{\cxt{coNP}}

\def\ptime{\cxt P}

\DeclareMathOperator\poly{poly}

\def\psp{\cxt{PSPACE}}

\def\st{\expandafter\hat}

\let\lgc\mathbf

\def\IPC{\lgc{IPC}}

\def\luk{\textbf{\textup\L}}
\def\I{{\bullet}}

\def\Q{\mathbb Q}
\def\Z{\mathbb Z}
\def\RR{\mathbb R}%\R clashes

\mathcode`\*="0203
\let\ob\overline

\mathchardef\mhyphen="2D

\def\noproof{\leavevmode\unskip\bme\vadjust{}\nobreak\hfill$\qed$\par}
\let\qed\Box
\newenvironment{Pf}[1][]
  {\par\noindent\textit{Proof\optpar{#1}:}\bme\ignorespaces}
  {\noproof\pagebreak[2]\vskip\medskipamount\ignorespacesafterend}
\def\qedhere{\relax\ifmmode\eqno\qed\expandafter\aftergroup
                   \else\noproof\fi\noqed}
\def\noqed{\let\noproof\relax}

\theoremstyle{plain}
\ifx\shortthm\undefined
\newtheorem{Thm}{Theorem}[section]
\else
\newtheorem{Thm}{Theorem}
\fi

\newtheorem{Cor}[Thm]{Corollary}
\newtheorem{Lem}[Thm]{Lemma}

\newtheorem{Cl}{Claim}[Thm]
\def\theCl{\arabic{Cl}}

\theorembodyfont\upshape

\newtheorem{Rem}[Thm]{Remark}

\newenvironment{Pf*}{\let\qed\qedCl\Pf}\endPf

\usepackage[reftex]{theoremref}

{\catcode`\^^I=13 \catcode`\^^M=13
\gdef\doalgo#1#2\end#{\hbox to\hsize{\hss \let^^I\qquad%
  \def\\^^M{\nobreak\hfil\break\vadjust{}\qquad}%
  \fboxsep1em \linenum0 %
  \fbox{\hsize#1\vbox{%
  \everypar{\advance\linenum1 %
      \hbox to1em{\hss$\scriptstyle\the\linenum$}\quad}%
  #2}}\hss}\end}}
\newcount\linenum

\def\key{\relax\ifmmode\expandafter\mathbf\else\expandafter\textbf\fi}
\def\allowhyphens{\nobreak\hskip0pt\relax}

\DeclareRobustCommand*\magiclparen{\ifmmode(\else\textup(\allowhyphens\fi}
\DeclareRobustCommand*\magicrparen{\ifmmode)\else\textup)\fi}
\let\lparen=(  \let\rparen=)
\def\magicparon{\catcode`\(\active\catcode`\)\active}
\def\magicparoff{\catcode`\(12 \catcode`\)12 }
\AtBeginDocument{\ifx\ifPreview\iftrue\else\magicparon\fi}
\magicparon
\let (=\magiclparen  \let )=\magicrparen

\ifx\enumup\undefined
  
\else
  
\fi

\def\optpar#1{\ifx\relax#1\relax\else\/ (#1)\fi}
\magicparoff

\def\cmb{\mathcode`\,"8000 }
\mathchardef\comma=\mathcode`\,
{\catcode`\,=\active \gdef,{\comma\penalty\relpenalty}}

\providecommand\dedic{}
\author{Emil Je\v r\'abek\dedic\\[\medskipamount]
Institute of Mathematics of the Academy of Sciences\\
\small \v Zitn\'a 25,
115\:67 Praha 1,
Czech Republic,
email: \texttt{jerabek@math.cas.cz}
}

\let\chull\relax
\DeclareMathOperator\chull{Conv}

\title{The complexity of admissible rules of \L ukasiewicz logic}

\begin{document}
\maketitle
\begin{abstract}
We investigate the computational complexity of admissibility of
inference rules in infinite-valued \L ukasiewicz propositional
logic ($\luk$). It was shown in \cite{ej:lukadm} that admissibility in
$\luk$ is checkable in $\psp$. We establish that this result is optimal,
i.e., admissible rules of $\luk$ are $\psp$-complete. In contrast,
derivable rules of $\luk$ are known to be $\conp$-complete.

\end{abstract}

\section{Introduction}\label{sec:introduction}
The concept of admissible rules was introduced by Lorenzen
\cite{loren}: a rule is admissible in a logical system if the set of
theorems (tautologies) of the logic is closed under instances of the
rule. In contrast to this, a rule is said to be derivable in a logic
if it belongs to its usual consequence relation. In classical logic,
derivable and admissible rules coincide (such logics are known as
structurally complete), but nonclassical logics typically have
nonderivable admissible rules, and often admissible rules exhibit much
more complicated structure than derivable rules.

Admissible rules are well understood for certain classes of transitive
modal and superintuitionistic logics. Admissibility in such logics was
investigated in a series of papers by Rybakov, culminating in the
monograph \cite{ryb:bk}. Another impetus was provided by the
characterization of unification and admissibility in terms of
projective formulas, introduced by Ghilardi \cite{ghilil,ghil}. This
incited work on bases of admissible rules including Iemhoff
\cite{iem:aripc,iem:arimed,iem:imed2} and Je\v r\'abek
\cite{ejadm,ej:indep}. Rybakov has recently studied admissible rules
in some temporal logics, see e.g.\ \cite{ryb:ldtl,ryb:ltlub}.

The computational complexity of admissibility of rules in modal and
superintuitionistic logics was investigated by Je\v r\'abek
\cite{ej:admcomp}. In particular, admissible rules of typical
transitive logics (e.g., $\IPC$, $\lgc{K4}$, $\lgc{S4}$, $\lgc{GL}$,
$\lgc{Grz}$) are $\cxt{coNEXP}$-complete, in contrast to derivable
rules of these logics, which are usually $\psp$-complete. (The
$\cxt{coNEXP}$-hardness part of the result holds for a quite wide class of logics,
including even $\conp$-logics of bounded depth such as
$\lgc{K4BD_3}$.) On the other hand, admissibility has the same
complexity as derivability in structurally complete and almost
structurally complete logics such as extensions of $\lgc{S4.3}$ (for
a nontrivial example of another kind, the $\{\to,\neg\}$-fragment of
$\IPC$ has $\psp$-complete admissibility problem by Cintula and
Metcalfe \cite{cin-met:intfrag}). Wolter and Zakharyaschev
\cite{wolt-zakh:undec} proved that unification and
admissibility in the extension of $\lgc K$ or $\lgc{K4}$ with the
universal modality is undecidable.

Admissible rules of \L ukasiewicz logic were investigated by Je\v
r\'abek \cite{ej:lukadm,ej:lukbas}. The main result of
\cite{ej:lukadm} is a description of a geometric criterion for
admissibility of multiple-conclusion rules in $\luk$, which in
particular implies that admissibility in $\luk$ (of single-conclusion
or multiple-conclusion rules, as well as the universal theory of free
$\M{MV}$-algebras) is computable in $\psp$. However, no nontrivial
lower bound on the complexity of admissibility in $\luk$ is given,
except that \L ukasiewicz tautologies are $\conp$-complete by Mundici
\cite{mun:luk}. In \cite{ej:lukbas}, an explicit basis of admissible
rules of $\luk$ is presented, and a description of admissibly
saturated formulas of $\luk$ is given. Recently, Marra and Spada
\cite{mar-spa} established that unification in $\luk$ is nullary
(i.e., of the worst possible type), and Cabrer \cite{cab:exact} proved
that admissibly saturated formulas in $\luk$ are exact.

The purpose of this paper is to show that the $\psp$ upper bound on
the complexity of admissibility in $\luk$ from \cite{ej:lukadm} is in
fact optimal: admissibility in $\luk$ is $\psp$-complete. The main
technical ingredient is a construction of a representation of the
configuration graph of a polynomial-space Turing machine by a rational
polyhedron which can be described by a polynomial-size \L ukasiewicz
formula. We also show an exponential lower bound on the length of
paths involved in the main criterion for admissibility in $\luk$ from
\cite{ej:lukadm} (matching an exponential upper bound given there).

The paper is organized as follows. In Section~\ref{sec:preliminaries}
we provide some background and fix the notation.
Section~\ref{sec:ar-luk} presents the criterion for admissibility in
$\luk$ from \cite{ej:lukadm} and provides an example where the
criterion requires exponentially long paths.
Section~\ref{sec:pspace-completeness} is devoted to the proof of our
main result, viz.\ $\psp$-completeness of admissibility in $\luk$.
Section~\ref{sec:conclusion} consists of concluding remarks.

\section{Preliminaries}\label{sec:preliminaries}
We assume the reader is familiar with basic notions from computational
complexity theory, such as Turing machines and the definitions of time
and space complexity. We recall that $\np$ is the class of languages
accepted by polynomial-time nondeterministic Turing machines, and
$\psp$ is the class of languages accepted by polynomial-space Turing
machines (whether deterministic or nondeterministic is immaterial here,
by Savitch's theorem). A language $L$ is $\psp$-complete if
$L\in\psp$, and every $\psp$-language is polynomial-time reducible to
$L$. The reader can consult e.g.\ Arora and Barak \cite{aro-bar} for
details and further background.

The \emph{standard $\M{MV}$-algebra} is the structure
$[0,1]_\luk=\p{[0,1],\cdot_\luk,\to_\luk,\min,\max,0,1}$ in the
signature $L_\luk=\p{\cdot,\to,\land,\lor,\bot,\top}$, where
$x\cdot_\luk y=\max\{0,x+y-1\}$ and $x\to_\luk y=\min\{1,1-x+y\}$. The
language of \L ukasiewicz logic ($\luk$) consists of propositional
formulas built freely from variables $x_i$, $i\in\omega$, and
connectives from $L_\luk$. (We will sometimes employ other letters,
such as $t,u,v$, for propositional variables.) A valuation is a
homomorphism $e$ from the free algebra of formulas into $[0,1]_\luk$.
A formula $\fii$ is an \emph{$\luk$-tautology} if $e(\fii)=1$
for every valuation $e$. A \emph{substitution} is an endomorphism on
the algebra of formulas. A substitution $\sigma$ is a \emph{unifier}
of a formula $\fii$ if $\sigma(\fii)$ is an $\luk$-tautology. A
\emph{rule} is an expression $\Gamma\ru\fii$, where $\Gamma$ is a
finite set of formulas. Such a rule is \emph{admissible} if every
common unifier of $\Gamma$ is also a unifier of $\fii$. More generally,
a \emph{multiple-conclusion rule} is an expression $\Gamma\ru\Delta$,
where $\Gamma,\Delta$ are finite sets of formulas; it is admissible if
every common unifier of $\Gamma$ is also a unifier of some formula
from $\Delta$. We write $\Gamma\adm_\luk\Delta$ if $\Gamma\ru\Delta$
is an admissible rule.

\emph{McNaughton's theorem} \cite{mcnau} states that a function
$\fii\colon[0,1]^m\to[0,1]$ is representable by a \L ukasiewicz
formula in $m$ variables if and only if it is a \emph{McNaughton
function}, i.e., a continuous piecewise linear (more precisely,
affine) function with integer coefficients. We will identify formulas
with their McNaughton functions when their syntactic shape is not
relevant. For any McNaughton function $\fii$, its \emph{truth set}
$t(\fii):=\fii^{-1}(1)$ is a \emph{rational polyhedron}: we can write
$t(\fii)=\bigcup_{i<k}C_i$, where each $C_i$ is a rational polytope,
i.e., the convex hull of a finite subset of $\Q^m$. Conversely, any
rational polyhedron $P\sset[0,1]^m$ equals $t(\fii)$ for some formula
$\fii$. We will write $t(\Gamma):=\bigcap_{\fii\in\Gamma}t(\fii)$, and
we denote the convex hull of a set $X\sset\RR^m$ by $\chull(X)$. We
have the following quantitative version of the easy implication in
McNaughton's theorem (see e.g.\ \cite{ej:lukadm}):
\begin{Lem}\th\label{lem:compolyt}
Let $\Gamma$ be a finite set of formulas in $m$ variables closed under
subformulas, and $n=\lh\Gamma$. For all $j<2^n$, $i<n$, and
$\fii\in\Gamma$, there are linear functions
$L_{j,i}$ and $L_{j,\fii}$ with integer coefficients and
$L^1$-norm at most $n$ such that the polytopes
$$C_j=\{x\in[0,1]^m:\forall i<n\,L_{j,i}(x)\ge0\}$$
satisfy
$$\bigcup_{j<2^n}C_j=[0,1]^m,$$
and
$$L_{j,\fii}(x)=\fii(x)$$
for each $x\in C_j$ and $\fii\in\Gamma$. Moreover, we can compute the
coefficients of $L_{j,i}$ and $L_{j,\fii}$ in polynomial time given
$\Gamma$ and $j$.
\noproof\end{Lem}
This also implies similar bounds on the expression of $t(\Gamma)$ as a
rational polyhedron.

\section{Admissible rules of \L ukasiewicz logic}\label{sec:ar-luk}
The following characterization of admissibility in $\luk$ was given in
\cite{ej:lukadm}. First, let us say that a set $X\sset\RR^m$ is
\emph{anchored} if its affine hull contains a \emph{lattice point}
(i.e., an element of $\Z^m$). Using efficient
computability of Herbrand's normal form, it can be seen that given a
sequence $x_1,\dots,x_n\in\Q^m$, it is polynomial-time decidable
whether $\{x_1,\dots,x_n\}$ is anchored.
\begin{Thm}[Je\v r\'abek \cite{ej:lukadm}]\th\label{thm:admdesc}
Let $\Gamma$ and $\Delta$ be finite sets of formulas in $m$ variables, and
let $\{C_j:j<r\}$ be a sequence of rational polytopes such that
$\bigcup_{j<r}C_j=t(\Gamma)$. The following are equivalent.
\begin{enumerate}
\item $\Gamma\nadm_\luk\Delta$.
\item There exists $a\in\{0,1\}^m\cap t(\Gamma)$ such that for
every $\psi\in\Delta$ there exists a sequence $\{j_i:i\le k\}$
of indices $j_i<r$ such that
\begin{enumerate}
\item $a\in C_{j_0}$,
\item $C_{j_i}$ is anchored for each $i\le k$,
\item $C_{j_i}\cap C_{j_{i+1}}\ne\nul$ for each $i<k$,
\item there exists $x\in C_{j_k}$ such that $\psi(x)<1$.
\noproof\end{enumerate}
\end{enumerate}
\end{Thm}
We can rephrase this in graph-theoretic language as follows. Given
$\Gamma$, consider the
decomposition  $t(\Gamma)=\bigcup_{j<r}C_j$,
$r\le2^n$, from \th\ref{lem:compolyt}. Let the \emph{polytope graph}
$G_\Gamma=\p{V_\Gamma,E_\Gamma}$ be the graph with vertex set
$V_\Gamma=\{0,\dots,r-1\}$ such that $j$ and $j'$ are connected by an edge in
$E_\Gamma$ iff $C_j\cap C_{j'}\ne\nul$. Let the \emph{anchored
polytope graph} $A_\Gamma$ be the induced
subgraph of $G_\Gamma$ consisting of vertices $j$ such that $C_j$ is
anchored. Let us call $j$ a \emph{lattice vertex} if
$C_j\cap\{0,1\}^m\ne\nul$, and $j$ is a \emph{counterexample} to a
formula $\psi$ if there exists $x\in C_j$ such that $\psi(x)<1$.
\begin{Cor}\th\label{cor:admgrf}
$\Gamma\nadm_\luk\Delta$ iff there exists a connected component of
$A_\Gamma$ containing a lattice vertex and a counterexample to $\psi$
for every $\psi\in\Delta$.
\noproof\end{Cor}
We also have:
\begin{Thm}[Je\v r\'abek \cite{ej:lukadm}]\th\label{thm:pspace}
$\adm_\luk$ is computable in $\psp$. \noproof\end{Thm} The original
proof of \th\ref{thm:pspace} in \cite{ej:lukadm} was a bit complicated
due to an effort to optimize the space requirements of the algorithm.
However, if we are not interested in a particular polynomial bound, we
can easily understand \th\ref{thm:pspace} as follows. Since we can
check in $\np$ whether a given polytope contains a lattice point or is
a counterexample to $\psi$ (the latter is even in $\ptime$, using
linear programming), \th\ref{cor:admgrf} reduces (non)admissibility in
$\luk$ to reachability in $A_\Gamma$. If an undirected graph is
explicitly given by a list of vertices and edges, reachability is
computable in logarithmic space (even deterministic, by a breakthrough
result of Reingold \cite{sl=l}; however, nondeterministic would do the
job for us). Instead of an input tape, the algorithm can be
implemented using oracle access to a black box which can tell whether
a given label denotes a valid vertex of the graph, and given two
vertices, whether they are connected by an edge. Now, our graph is
exponentially large, which blows up the complexity from logarithmic to
polynomial space. The whole algorithm is $\psp$ provided we can
simulate the input oracle in polynomial space as well. In fact, we can
do it in $\np$: given $j$, we can compute the linear functions
defining the polytope $C_j$; then we can check in $\np$ whether it is
anchored, and given two such polytopes, we can check whether they
intersect.

It should be clear from this description that the only obstacle
preventing us from computing $\adm_\luk$ more efficiently is that the
path connecting in $A_\Gamma$ a counterexample to $\psi$ to a lattice
vertex may be exponentially long. For example, it is not difficult to
see that if we could always find such a path of polynomial length, we
could test $\nadm_\luk$ in $\np$. Thus, if we intend to prove that
$\adm_\luk$ is $\psp$-complete, we had better make sure that
there are cases where the distance from any counterexample to $\psi$
to any lattice vertex is exponentially long.

The construction in the proof of our main result will indeed have this
property (when applied to an exponential-time $\psp$ algorithm).
However, we decided to also include a simpler direct construction,
since it illustrates more transparently the motivation behind the
general case, which may help the reader in understanding the
underlying idea. \th\ref{prop:exppath} and its proof are not needed
for our main result, hence a reader who wants to get straight to the
point may safely skip to the next section.

\begin{Thm}\th\label{prop:exppath}
Given $m$, we can construct in time $\poly(m)$ formulas
$\fii_m,\psi_m$ of size $O(m^2)$ in $m$ variables such that
$\fii_m\nadm_\luk\psi_m$, but every sequence $\{j_i:i\le k\}$ as in
\th\ref{thm:admdesc} must have length $k=\Omega(2^m)$.
\end{Thm}
\begin{Pf}
Let $G_m=\p{V_m,E_m}$ be the $m$-dimensional hypercube graph: i.e.,
$V_m=\pw m$ (where we use the set-theoretical identity
$m=\{0,\dots,m-1\}$ to simplify the notation), 
and $\p{u,v}\in E_m$ iff $\lh{u\sdif v}=1$, where $\sdif$
denotes symmetric difference. We will define an exponentially long
path $P_m$ in $G_m$, and embed $G_m$ in $[0,1]^m$ in such a way that
$P_m$ is represented by the graph $A_\fii$ for a polynomial-size formula $\fii$.

The path $P_m=\p{v_{m,0},\dots,v_{m,2^m-1}}$ will be a Hamiltonian
path in $G_m$ starting at the vertex $v_{m,0}=\nul$, and we define it
inductively as follows: $P_0$ is the trivial one-vertex path in $G_0$.
If $P_m$ was already constructed, we define $P_{m+1}$ by taking two
copies of $P_m$, one in each of the hyperplanes $\{v\sset m+1:m\notin
v\}$ and $\{v\sset m+1:m\in v\}$, and joining them by an edge
connecting the two copies of the far end-point of $P_m$. That is,
$$P_{m+1}=\p{v_{m,0},\dots,v_{m,2^m-1},v_{m,2^m-1}\cup\{m\},\dots,v_{m,0}\cup\{m\}}.$$
We will actually need a more explicit description of the edges
belonging to $P_m$. First, since $v_{m,0}=\nul$ for every $m$, the
other end-point of $P_m$ is $v_{m,2^m-1}=\{m-1\}$ for $m>0$. Then it
is easy to show by induction on $m$ that every vertex $v\in V_m$ is
connected in $P_m$ to
\begin{itemize}
\item $v\sdif\{0\}$, and
\item $v\sdif\{\min(v)+1\}$ if possible (i.e., if $v\ne\nul,\{m-1\}$).
\end{itemize}
We can identify each $v\sset m$ with the binary string describing its
characteristic function. That is, we make $V_m=\{0,1\}^m$, and  then
$P_m$ consists of the following edges, where we denote concatenation
by juxtaposition:
\begin{itemize}
\item $0w$---$1w$, for $w\in\{0,1\}^{m-1}$,
\item $0^k10w$---$0^k11w$, for $k<m-1$, $w\in\{0,1\}^{m-k-2}$.
\end{itemize}
The end-points of $P_m$ are $0^m$ and $0^{m-1}1$. By abuse of
language, we will denote the set of edges of $P_m$ as $P_m$.

We now construct a representation of $G_m$ in $[0,1]^m$. Put
$B_0=[0,1/5]$, $B_1=[3/5,4/5]$, and $B=[0,4/5]$. We represent a vertex
$v\in\{0,1\}^m$ by the polytope
$$B_v=\prod_{i<m}B_{v_i}.$$
If $e=\{v,w\}\in E_m$, let $j<m$ be the unique position such that
$v_j\ne w_j$. We represent $e$ by the polytope
$$C_e=\prod_{i\ne j}B_{v_i}\times B,$$
where the $B$ is supposed to go to the $j$th position in the product.
Let $$C=\bigcup_{e\in P_m}C_e.$$
The following properties are easy to verify:
\pagebreak[2]
\begin{Cl}\th\label{lem:toyemb}
\ \begin{enumerate}
\item Each $B_v$ and $C_e$ is an anchored rational polytope.
\item $B_v$ are pairwise disjoint.
\item If $v\in e$, then $B_v\sset C_e$, otherwise $B_v\cap C_e=\nul$.
\item $C_e$ are pairwise disjoint, except that $C_e\cap
C_{e'}=B_v$ when $e\cap e'=\{v\}$.
\item $B_v$ contains a lattice point iff $v=0^m$. $C_e$ contains a
lattice point iff\/ $0^m\in e$.
\item $C$ is connected. If $v\ne0^m,0^{m-1}1,$ then $C\bez B_v$ is disconnected, and its 
two connected components correspond to the two subpaths of $P_m$ on
either side of $v$.
\end{enumerate}
\end{Cl}
The key property is that even though
there are exponentially many edges in $P_m$, we can write $C$ in
another way using only polynomially many operations, because of the
highly uniform way in which $P_m$ can be described. Indeed,
$$C=(B\times B_*^{m-1})\cup
  \bigcup_{k<m-1}(B_0^k\times B_1\times B\times B_*^{m-k-2}),$$
where $B_*=B_0\cup B_1$. Fix formulas $\beta_0$, $\beta_1$, $\beta$ in
one variable such that $t(\beta_i)=B_i$, $t(\beta)=B$, and put
$\beta_*=\beta_0\lor\beta_1$. Then we have $C=t(\fii_m)$, where
$$\fii_m=\Bigl(\beta(x_0)\land\ET_{i=1}^{m-1}\beta_*(x_i)\Bigr)\lor
  \LOR_{k<m-1}\Bigl(\ET_{i<k}\beta_0(x_i)\land\beta_1(x_k)
      \land\beta(x_{k+1})\land\ET_{i=k+2}^{m-1}\beta_*(x_i)\Bigr).$$
Notice that $\lh{\fii_m}=O(m^2)$.
Let $\delta_i$ be fixed formulas in one variable such that
$t(\delta_i)=[0,1]\bez\Int(B_i)$, and put
$$\psi_m=\LOR_{i<m-1}\delta_0(x_i)\lor\delta_1(x_{m-1}),$$
so that
$$t(\psi_m)=D:=[0,1]^m\bez\Int(B_{0^{m-1}1}).$$
Since $C$ is a connected union of anchored polytopes, contains a
lattice point $\vec 0$, and a counterexample to $\psi_m$, we have
$$\fii_m\nadm_\luk\psi_m.$$
On the other hand, if we write $t(\fii)$ as $\bigcup_{e\in
P_m}C_e$, then it follows from \th\ref{lem:toyemb} that the
only path in $A_\fii$ connecting a lattice vertex to a counterexample
to $\psi_m$ traces $P_m$ all the way from one end to the other end,
hence it has length $2^m-1$.

A subtle issue (which will not arise in the $\psp$-completeness
proof below) is that in principle it may be possible to write $C$ as
a union of polytopes $\bigcup_{i<r}C'_i$ in a different way so that
there is a shorter path from a lattice vertex to a counterexample to
$\psi_m$. However, we have:
\pagebreak[2]
\begin{Cl}\th\label{lem:toyconvex}
Any convex subset of $C$ intersects at most two $B_v$.
\end{Cl}
\begin{figure}
\begin{center}
\includegraphics{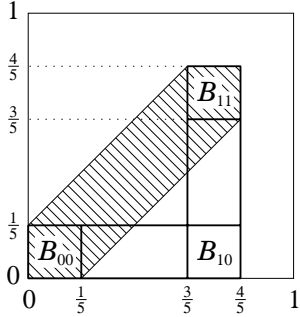}
\end{center}
\caption{The convex hull of $B_{00}\cup B_{11}$ is disjoint from $B_{10}$}
\label{fig:hull}
\end{figure}
\begin{Pf*}
Let $X\sset C$ be convex. If $x\in X\cap B_u$ and $y\in X\cap B_w$,
the line segment $\chull(x,y)$ is included in $X\sset C$ and it is
connected,
hence by \th\ref{lem:toyemb} it hits $B_v$ for every $v$ lying on the
subpath of $P_m$ joining $u$ to $w$. Thus, if we assume for
contradiction that $X$ intersects three or more $B_v$, we can find
$u,v,w$ such that $\{u,v\},\{v,w\}\in P_m$, $x\in X\cap B_u$, $y\in
X\cap B_w$, $\chull(x,y)\cap B_v\ne\nul$. Let $i\ne j$ be the unique
coordinates such that $u_i\ne v_i$ and $v_j\ne w_j$, and let $\pi$ be
the projection to the $i$th and $j$th coordinates. Then
$\pi(B_u)=B_{u_iu_j}$ and similarly for $B_v,B_w$, and $\pi$ preserves
convex hulls, hence there exist $u',v',w'\in\{0,1\}^2$ such that
$\{u',v'\},\{v',w'\}\in P_2$, and $\chull(B_{u'}\cup B_{w'})\cap
B_{v'}\ne\nul$. However, this is easily seen to be false, see
Figure~\ref{fig:hull}.
\end{Pf*}
By \th\ref{lem:toyemb}, removing any $B_v$ from $C$ disconnects the
unique lattice point $\vec0$ from $C\bez D$, hence any path using the
$C'_i$ witnessing $\fii_m\nadm_\luk\psi_m$ as in \th\ref{thm:admdesc}
must intersect every $B_v$. By \th\ref{lem:toyconvex}, such a path has to
have length at least $2^{m-1}$.
\end{Pf}
\section{$\psp$-completeness}\label{sec:pspace-completeness}
We will use an idea similar to the proof of \th\ref{prop:exppath}
to simulate a computation of a
polynomial-space Turing machine. In a nutshell, we will embed in
$[0,1]^m$ the configuration graph of the machine. (This subsumes the
ability to create exponentially long paths as a polynomial-space
computation may take exponential time.) In order to get a description
of the graph by a polynomial-size formula, we will exploit the
locality of Turing machines: the behaviour of the machine in a
particular configuration is determined by a constant-size subset of
the configuration, and anything outside this subset is passed
unchanged to the next step.

In order to simplify the construction, we will not simulate completely
general polynomial-space Turing machines, but we will first reduce to
a special case that is more manageable. Let us say that a
deterministic Turing machine $M$ is in a \emph{normal form} if it has
the following properties. $M$ has a single tape with alphabet
$\Sigma=\{0,1\}$ (using no extra blank symbol) which serves both as
the input tape and as a work tape. $M$ has states with labels from
$Q=\{0,\dots,s\}$, $s\ge1$, where $0$ is the initial state, and $1$ is
the unique accepting state. There is no rejecting state, on
non-accepted inputs $M$ eventually enters an infinite loop. The tape
head moves left or right in every step. Let $T\colon Q\times\Sigma\to
Q\times\Sigma\times\{1,-1\}$ be the transition function of $M$ (i.e.,
when $M$ is in state $q$ with the tape head in position $h$ reading symbol
$x\in\Sigma$, and $T(q,x)=\p{r,y,d}$, then $M$ writes $y$ to the tape,
moves head to position $h+d$, and enters state $r$). We require
$T(1,x)=\p{1,y,d}$; i.e., $T$ is defined in such a way that once $M$
enters the accepting state, it can never leave it. (This is only a
formal technical requirement, as after entering the accepting state
$M$ is supposed to stop anyway. However, it will be convenient for our
simulation to pretend that the machine continues to work in order to
reduce the number of exceptions.) On an input $w\in\{0,1\}^n$, $M$
starts with head at position $0$ of the tape and $w=w_0\dots w_{n-1}$
written at positions $0,\dots,n-1$ of the tape. A \emph{normal run} of
$M$ on input of length $n$ is a computation during which $M$ does not
attempt to access positions $-1$ or $n$ of the tape (which in
particular implies that it is confined to space $n$). We consider
acceptance by $M$ as a promise problem, whose positive instances are
inputs accepted by a normal run of $M$, and negative instances are
inputs that make $M$ enter an infinite normal run avoiding the
accepting state.
\begin{Lem}\th\label{lem:normal}
Every $L\in\psp$ is polynomial-time reducible to the acceptance
problem of a Turing machine in normal form.
\end{Lem}
\begin{Pf}
Let $L\sset\Sigma_0^*$, and let $M_1$ be a deterministic Turing
machine accepting $L$ in space $p(n)\ge n$ using $k$ work tapes (along
with the input tape) with alphabet
$\Sigma_1\Sset\Sigma_0\cup\{\epsilon\}$, where $\epsilon$ is the blank
symbol, and $p$ is a polynomial.
Let $\Sigma'_1=\{a':a\in\Sigma_1\}$ be a disjoint copy of $\Sigma_1$,
and $\dia\notin\Sigma_1\cup\Sigma'_1$ an auxiliary symbol. We can
represent a configuration $c$ of $M_1$ by the string
$$
\tilde c=\dia\tilde a_0^0\tilde a_1^0\dots\tilde a_{p(n)-1}^0
\dia\tilde a_0^1\tilde a_1^1\dots\tilde a_{p(n)-1}^1
\dia\cdots
\dia\tilde a_0^k\tilde a_1^k\dots\tilde a_{p(n)-1}^k\dia,
$$
where $a_i^j$ is the $i$th symbol on the $j$th tape (the input tape
being the $0$th tape), and $\tilde a_i^j=(a_i^j)'$ if the head of tape
$j$ is on position $i$, $\tilde a_i^j=a_i^j$ otherwise. We can
simulate easily the computation of $M_1$ by a single-tape Turing
machine $M_2$ with alphabet $\Sigma_2=\Sigma_1\cup\Sigma'_1\cup\{\dia\}$
operating with the representations $\tilde c$ of configurations of
$M_1$ in such a way that $M_2$ never attempts to move past the first
or last $\dia$ delimiters. Choose $d\in\omega$ and pairwise distinct
$\ob a\in\{0,1\}^d$ for each $a\in\Sigma_2$. We can simulate $M_2$ by
a machine $M$ in normal form by translating each symbol $a$ of the
simulated tape of $M_2$ with the sequence $\ob a$ of $d$ binary
symbols. A run of $M$ is normal whenever it starts with the tape
containing the translation of a
valid representation $\tilde c$ of a configuration of $M_1$. Then $L$
is reducible to the acceptance problem of $M$ via the polynomial-time
function $f(x)$ which computes the translation of $\tilde c$, where
$c$ is the initial configuration of $M_1$ on input $x$.
\end{Pf}
\begin{Thm}\th\label{thm:complete}
Admissibility of either single-conclusion or multiple-conclusion rules
in $\luk$ is $\psp$-complete.
\end{Thm}
\begin{Pf}
That ${\adm}_\luk\in\psp$ was established in
\cite{ej:lukadm}, hence it suffices to show that non-admissibility of
single-conclusion rules in $\luk$ is $\psp$-hard. Given a
$\psp$ language $L$, let $f$ be a
polynomial-time function and $M$ a Turing machine in normal form such
that $x\in L$ iff $M$ accepts $f(x)$, and the run of $M$ on any
$w=f(x)$ is normal.

Let $n$ be given. A configuration of $M$ is a sequence
$c=\p{q,h,x_0,\dots,x_{n-1}}$, where $q\in Q=\{0,\dots,s\}$ is the
current state,
$h<n$ is the position of the head, and $x_0,\dots,x_{n-1}$ is the
content of the tape. Put $I_0=[1/5,2/5]$, $I_1=[3/5,4/5]$,
$I_*=I_0\cup I_1$, $I_\I=[2/5,3/5]$, $I'_0=[0,1/5]$, $I'_1=[4/5,1]$,
$J_q=[(2q+1)/(2s+3),(2q+2)/(2s+3)]$ for $q\le s$, $J_*=\bigcup_{q\le s}J_q$,
$J=[1/(2s+3),(2s+2)/(2s+3)]=\chull(J_*)$, $J'_0=[0,1/(2s+3)]$ (cf.\
Figure~\ref{fig:int}).
\begin{figure}
\begin{center}
\includegraphics{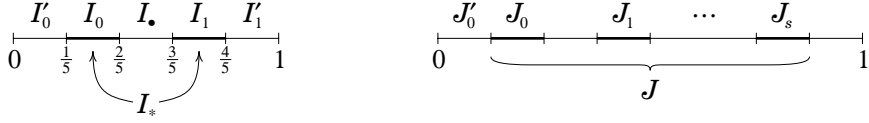}
\end{center}
\caption{The layout of auxiliary intervals}
\label{fig:int}
\end{figure}
We represent a
configuration $c=\p{q,h,x_0,\dots,x_{n-1}}$ by the polytope
$$H_c=J_q\times\prod_{i<n}I_{\delta_{h,i}}\times\prod_{i<n}I_{x_i}
\sset[0,1]^{2n+1},$$
where $\delta_{h,i}$ is Kronecker's delta. We
represent the input $w=f(x)$ of length $n$ by
$$F_w=J'_0\times\prod_{i<n}I'_{\delta_{0,i}}
  \times\prod_{i<n}I'_{w_i}.$$
Acceptance by $M$ will be represented by (the complement of) the
polyhedron
$$B=[0,1]^{2n+1}\bez\Int(J_1\times I_*^{2n}).$$
Finally, we have to find a representation for transition edges. For
any configuration $c$, let $\sigma(c)$ be its successor configuration
(which is unique, as $M$ is deterministic). We will construct a
polyhedron $E_c$ representing an edge connecting $c$ to $\sigma(c)$
as follows.
\begin{Cl}\th\label{cl:gadget}
For every $q\in Q$ and $x\in\{0,1\}$, we can choose a
rational polyhedron $C_{q,x}\sset[0,1]^4$ with the following
properties, where $T(q,x)=\p{r,y,d}$ is the transition function of $M$:
\begin{enumerate}
\item $C_{q,x}$ is connected, and it is a finite union of polytopes of
dimension $4$.
\item $C_{q,x}$ intersects
$J_q\times\{\p{3/5,2/5,(2+x)/5}\}$ and
$J_r\times\{\p{2/5,3/5,(2+y)/5}\}$.
\item $C_{q,x}$ is included in $J\times I_\I^3$, and more
precisely, in
$$(J_q\times\{\p{3/5,2/5,(2+x)/5}\})
 \cup(J_r\times\{\p{2/5,3/5,(2+y)/5}\})
 \cup(J\times\Int(I_\I)^3).$$
\item The sets $C_{q,x}$ are pairwise disjoint.
\end{enumerate}
\end{Cl}
\begin{Pf*}
The reader may well take it on faith that there is room enough in the
$4$-dimensional space to embed a finite collection of edges, but for
definiteness, we can construct $C_{q,x}$ explicitly as follows. Let
us enumerate $Q\times\{0,1\}=\{\p{q_i,x_i}:i<m\}$ (hence $m=2(s+1)$),
and put $\p{r_i,y_i,d_i}=T(q_i,x_i)$.
Denote $[a\pm\ep]=[a-\ep,a+\ep]$ and $c(t,x,y)=(1-t)x+ty$. We put
$z_{q,i}=c((1+i)/(2m+1),\min(J_q),\max(J_q))$, $\ob
z_{q,i}=z_{q,m+i}$, $h_i=c((1+i)/(m+1),2/5,3/5)$. Let $C'_{q_i,x_i}$
be the broken line with end-points $\p{z_{q_i,i},3/5,2/5,(2+x_i)/5}$,
$\p{z_{q_i,i},1/2,1/2,h_i}$, $\p{\ob z_{r_i,i},1/2,1/2,h_i}$, $\p{\ob
z_{r_i,i},2/5,3/5,(2+y_i)/5}$. Then $C'_{q_i,x_i}$ satisfies all the
requirements above except that it has only dimension $1$. Let $\ep>0$,
$\ep\in\Q$ be such that the $L^\infty$-distance of $C'_{q_i,x_i}$ and
$C'_{q_{i'},x_{i'}}$ is at least $3\ep$ for each $i\ne i'$. We can
define $C_{q_i,x_i}$ to be the union of the following three polytopes:
\begin{enumerate}
\item The convex hull of $\p{z_{q_i,i},3/5,2/5,(2+x_i)/5}$ and
$[z_{q_i,i}\pm\ep]\times[1/2\pm\ep]^2\times[h_i\pm\ep]$,
\item $[z_{q_i,i},\ob z_{r_i,i}]\times[1/2\pm\ep]^2\times[h_i\pm\ep]$,
\item The convex hull of $\p{\ob z_{r_i,i},2/5,3/5,(2+y_i)/5}$ and
$[\ob z_{r_i,i}\pm\ep]\times[1/2\pm\ep]^2\times[h_i\pm\ep]$.
\end{enumerate}
Notice that $C_{q_i,x_i}$ is contained within the closed
$\ep$-neighbourhood of $C'_{q_i,x_i}$ (in the $L^\infty$-norm). Then
it is easy to see that $C_{q_i,x_i}$ satisfies all our requirements.
\end{Pf*}
Given a configuration $c=\p{q,h,x_0,\dots,x_{n-1}}$, let
$T(q,x_h)=\p{r,y,d}$, so that
$$\sigma(c)=\p{r,h+d,x_0,\dots,x_{h-1},y,x_{h+1},\dots,x_{n-1}}.$$
We define
$$E_c=C_{q,x_h}\times\prod_{i\ne h,h+d}I_0\times\prod_{i\ne h}I_{x_i},$$
where the four coordinates of $C_{q,x_h}$ are supposed to go to the
$0$th, $(h+1)$st, $(h+d+1)$st, and $(h+n+1)$st coordinates in the
product; that is, more precisely,
\begin{multline}\label{eq:ec}
E_c=\{\p{t,u_0,\dots,u_{n-1},v_0,\dots,v_{n-1}}:
  u_i\in I_0\ (i\ne h,h+d),\\ v_i\in I_{x_i}\ (i\ne h),
  \p{t,u_h,u_{h+d},v_h}\in C_{q,x_h}\}
\end{multline}
(cf.\ the definition of $H_c$).
We put $H=\bigcup_cH_c$, $E=\bigcup_cE_c$, $A_w=H\cup E\cup F_w$.
Notice that we have
\begin{equation}\label{eq:her}
\begin{aligned}
H&=J_*\times\bigcup_{h<n}(I_0^h\times I_1\times I_0^{n-h-1})
      \times I_*^n,\\
E&=\bigcup_{q,h,x}(C_{q,x}\times I_0^{n-2}\times I_*^{n-1}),\\
B&=\left(([0,1]\bez\Int(J_1))\times[0,1]^{2n}\right)\cup
  \bigcup_{i=1}^{2n}\left([0,1]^i\times([0,1]\bez\Int(I_*))
             \times[0,1]^{2n-i}\right),
\end{aligned}
\end{equation}
where the products in $E$ have coordinates permuted as in the
definition of $E_c$ above.
\pagebreak[2]
\begin{Cl}\th\label{cl:prop}
\ \begin{enumerate}
\item\label{item:geom} $H_c$ and $F_w$ are full-dimensional (hence anchored) polytopes.
$E_c$ is a connected finite union of full-dimensional polytopes.
\item\label{item:lat} There is no lattice point in $H\cup E$, and there is one in
$F_w$. $F_w$ is disjoint from $E$, and it intersects $H_c$ iff $c$ is
the initial configuration $\p{0,0,w}$.
\item\label{item:bdis} $H_c$ are pairwise disjoint.
\item\label{item:be} $E_c$ intersects $H_d$ iff $d=c$ or $d=\sigma(c)$.
\item\label{item:edis} $E_c\bez H$ are pairwise disjoint.
\item\label{item:r} $B\Sset E\cup F_w$. $B$ includes $H_c$ iff $c$ is not an
accepting configuration.
\item\label{item:acc} The connected component of $A_w$ containing $F_w$
is included in $B$ if and only if $M$ does not accept $w$.
\end{enumerate}
\end{Cl}
\begin{Pf*}
\eqref{item:geom}, \eqref{item:lat}, \eqref{item:bdis}, and
\eqref{item:r} are immediate from the definition.

\eqref{item:be}: Let $c=\p{q,h,x_0,\dots,x_{n-1}}$. $C_{q,x_h}$
intersects $J_q\times\{\p{3/5,2/5,(2+x_h)/5}\}\sset J_q\times I_1\times
I_0\times I_{x_h}$, hence $E_c$ intersects $H_c$. Similarly,
$C_{q,x_h}$ intersects $J_r\times I_0\times I_1\times I_y$, where
$\p{r,y,d}=T(q,x_h)$, hence $E_c$ intersects $H_{\sigma(c)}$. The
remaining part of $C_{q,x_h}$ is contained in $J\times\Int(I_\I)^3$,
and as $\Int(I_\I)\cap I_*=\nul$, the corresponding part of $E_c$ is
disjoint from $H$.

\eqref{item:edis}: By the proof of \eqref{item:be}, $E_c\bez H$
corresponds to the part of $C_{q,x_h}$ included in
$J\times\Int(I_\I)^3$. Let $c'=\p{q',h',x'_0,\dots,x'_{n-1}}$,
$T(q',x'_{h'})=\p{r',y',d'}$ be such that $E_c\cap E_{c'}\nsset H$. If
$h\ne h'$, the projection of $E_{c'}$ to the $v_h$-coordinate (using
the notation of \eqref{eq:ec}) is included in $I_*$, whereas $E_c\bez
H$ projects to the disjoint interval $\Int(I_\I)$, a contradiction.
Thus $h=h'$. If $d\ne d'$, then similarly the projections of $E_c\bez
H$ and $E_{c'}$ to the $u_{h+d}$-coordinate are included in
$\Int(I_\I)$ and $I_0$, respectively, hence we may assume $d=d'$. If
$x_i\ne x'_i$ for some $i\ne h$, then the projections of $E_c$ and
$E_{c'}$ to the $v_i$-coordinate are $I_{x_i}$ and $I_{x'_i}$.
Finally, if $x_i=x'_i$ for all $i\ne h$, then $E_c=C_{q,x_h}\times X$
and $E_{c'}=C_{q',x'_h}\times X$ for a certain set $X$, up to a
permutation of coordinates (the same one for both). Since the sets
$C_{q,x}$ are pairwise disjoint, we must have $q=q'$ and $x_h=x'_h$,
i.e., $c=c'$.

\eqref{item:acc}: Assume that the component is not included in $B$.
There exists a sequence $P_0,\dots,P_r$ of polyhedrons such that
$P_0=F_w$, each $P_i$ for $i>0$ is $H_c$ or $E_c$, $P_i\cap
P_{i+1}\ne\nul$, and $P_r\nsset B$. By \eqref{item:r}, $P_r=H_c$ for
some accepting configuration $c$. By \eqref{item:lat},
$P_1=H_{0,0,w}$. By \eqref{item:edis}, we may assume that no two $E_c$
are adjacent in the sequence. By \eqref{item:be}, this implies that
$E_c$ can only be adjacent to $H_c$ and $H_{\sigma(c)}$. By
\eqref{item:bdis}, no two $H_c$ are adjacent. Summing up, there exists
a sequence $c_0,\dots,c_p$ of pairwise distinct configurations such
that $c_0=\p{0,0,w}$ is the initial configuration, $c_p$ is an
accepting configuration, and for each $i<p$, $c_{i+1}=\sigma(c_i)$ or
$c_i=\sigma(c_{i+1})$. However, if $c_i=\sigma(c_{i+1})$ and
$c_{i+2}=\sigma(c_{i+1})$, then $c_i=c_{i+2}$, and we can delete
$c_{i+1}$ and $c_{i+2}$ from the sequence. Thus, we can assume that
there exists $j\le p$ such that $c_{i+1}=\sigma(c_i)$ for all $i<j$, and
$c_i=\sigma(c_{i+1})$ for all $i\ge j$. Since $c_p$ is an accepting
configuration and successors of accepting configurations are
again accepting, $c_j$ is also an accepting configuration, hence $M$ accepts
$w$.

Conversely, if $c_0,\dots,c_p$ is the sequence of configurations of
$M$ during an accepting computation on $w$, then the sequence
$F_w,H_{c_0},E_{c_0},H_{c_1},\dots,E_{c_{p-1}},H_{c_p}$ witnesses that
$F_w$ is in $A_w$ connected to the complement of $B$.
\end{Pf*}
We now express $A_w$ and $B$ by propositional formulas (using
variables $\cmb t,u_0,\dots,u_{n-1},v_0,\dots,v_{n-1}$ in the same
fashion as in \eqref{eq:ec}). Let
$\cmb\iota_0,\iota_1,\iota_*,\iota'_0,\iota'_1,\ob\iota_*,\zeta_*,\zeta'_0,\ob\zeta_1$
be formulas in one variable whose truth sets are $\cmb
I_0,I_1,I_*,I'_0,I'_1,{[0,1]}\bez\Int(I_*),J_*,J'_0,{[0,1]}\bez\Int(J_1)$,
respectively, and for any $q\le s$ and $x\in\{0,1\}$, let
$\gamma_{q,x}$ be a formula in four variables whose truth set is
$C_{q,x}$. Notice that these formulas only depend on $M$ and not on
$n$ or $w$, hence they are fixed constant-size formulas. Then we put
\begin{align*}
\eta_n&=\zeta_*(t)\land
  \LOR_{h<n}\ET_{i<n}\iota_{\delta_{h,i}}(u_i)\land\ET_{i<n}\iota_*(v_i),\\
\ep_n&=\LOR_{q,x,h,d}
   \Bigl(\gamma_{q,x}(t,u_h,u_{h+d},v_h)
      \land\ET_{i\ne h,h+d}\iota_0(u_i)
      \land\ET_{i\ne h}\iota_*(v_i)\Bigr),\\
\fii_w&=\zeta'_0(t)\land\iota'_1(u_0)
  \land\ET_{i=1}^{n-1}\iota'_0(u_i)
  \land\ET_{i<n}\iota'_{w_i}(v_i),\\
\alpha_w&=\eta_n\lor\ep_n\lor\fii_w,\\
\beta_n&=\ob\zeta_1(t)\lor
  \LOR_{i<n}(\ob\iota_*(u_i)\lor\ob\iota_*(v_i)),
\end{align*}
where the disjunction in $\ep_n$ is taken over all $q\le s$,
$x\in\{0,1\}$, $d\in\{1,-1\}$, and $h<n$ such that 
$T(q,x)=\p{r,y,d}$ and $0\le h+d<n$. It follows from \eqref{eq:her}
that $t(\eta_n)=H$, $t(\ep_n)=E$, $t(\fii_w)=F_w$, $t(\alpha_w)=A_w$,
and $t(\beta_n)=B$, hence using \th\ref{cl:prop} and \th\ref{thm:admdesc},
$$\alpha_w\nadm_\luk\beta_n\iff M\text{ accepts }w.$$
We have $\lh{\alpha_w}=O(n^2)$ and $\lh{\beta_n}=O(n)$, and it is easy
to see that $\alpha_w$ and $\beta_n$ are polynomial-time (or even
log-space) computable given $w$, hence
$$x\in L\iff\alpha_{f(x)}\nadm_\luk\beta_{\lh{f(x)}}$$
provides a polynomial-time reduction of $L$ to $\nadm_\luk$.
\end{Pf}
\begin{Rem}
It follows from \th\ref{thm:complete} that the quasi-equational
theory of free $\M{MV}$-algebras is $\psp$-hard. Since
the universal theory of free $\M{MV}$-algebras was shown to be in
$\psp$ in \cite{ej:lukadm}, both these theories are
$\psp$-complete.
\end{Rem}
\section{Conclusion}\label{sec:conclusion}
We have settled the computational complexity of admissibility in
$\luk$ by showing its $\psp$-completeness. One consequence is that the
algorithm for admissibility given in \cite{ej:lukadm} cannot be
significantly improved. Moreover, it confirms the intuition suggested
by the criterion from \cite{ej:lukadm} that admissibility in $\luk$ is
best viewed in terms of undirected reachability in the anchored polytope graph,
at least in the sense that it leads to the right complexity estimate
of the problem. It is also worth mentioning that similarly to the case
of natural transitive modal logic and intuitionistic logic, the
admissibility problem in $\luk$ turns out to be more complex than the
derivability problem (assuming $\np\ne\psp$).

Our result resolves Problem~5.2 from \cite{ej:lukadm}. We remark that
Problem~5.1 is also essentially solved: Marra and Spada \cite{mar-spa}
proved the unification type of $\luk$ to be nullary, which also shows
that some formulas cannot have projective approximations,
despite that all formulas have admissibly saturated approximations by
\cite{ej:lukbas}. The description of projective formulas in $\luk$
remains an intriguing open problem (some results in this direction
have been obtained by Cabrer and Mundici \cite{cab-mun}), nevertheless, in
view of the nonexistence of projective approximations, it is not
directly relevant to admissibility; a question more to the point is a
characterization of admissibly saturated formulas, which is
satisfactorily resolved by \cite{ej:lukbas,cab:exact}. Leaving
admissibility aside, an interesting related problem is to get a better
understanding of unification in $\luk$. For instance, despite its
nullary type, it is conceivable that one can describe (infinite)
complete sets of unifiers in some transparent algorithmic way.

\subsection*{Funding}
This work was supported by Institutional Research Plan
AV0Z10190503, grant IAA100190902 of GA AV \v CR, project 1M0545 of
M\v SMT \v CR, and a grant from the John Templeton Foundation.

\subsection*{Acknowledgements}
I would like to thank the referee for useful suggestions.

\bibliographystyle{mybib}
\bibliography{lukcxt}
\end{document}